\documentclass{elsart}
\usepackage{graphicx}
\usepackage{bbm}

\makeatletter
\def\lesssim{\mathrel{\mathpalette\vereq<}}
\def\vereq#1#2{\lower3pt\vbox{\baselineskip1.5pt \lineskip1.5pt
\ialign{$\m@th#1\hfill##\hfil$\crcr#2\crcr\sim\crcr}}}
\def\gtrsim{\mathrel{\mathpalette\vereq>}}
\makeatother

\begin{document}
\runauthor{Baur, Hencken, Aste, Trautmann, Klein}
\begin{frontmatter}
\title{
Multiphoton Exchange Processes in Ultraperipheral Relativistic
Heavy Ion Collisions
}
\author[FZJ]{Gerhard Baur}
\author[UBA]{Kai Hencken}
\author[UBA]{Andreas Aste}
\author[UBA]{Dirk Trautmann}
\author[LBNL]{Spencer R. Klein}

\address[FZJ]{Forschungszentrum J\"ulich, 52425 J\"ulich, Germany}
\address[UBA]{Universit\"at Basel, 4056 Basel, Switzerland}
\address[LBNL]{Lawrence Berkeley National Laboratory, Berkeley, 94720 CA, USA}

\begin{abstract}
The very strong electromagnetic fields present in ultraperipheral
relativistic heavy ion collisions lead to important higher order
effects of the electromagnetic interaction.  These multiphoton
exchange processes are studied using perturbation theory and the
sudden or Glauber approximation.  In many important cases, the
multi-photon amplitudes factorize into independent single-photon
amplitudes.  These amplitudes have a common impact parameter vector,
which induces correlations between the amplitudes.
Impact-parameter dependent equivalent-photon spectra for simultaneous
excitation are calculated, as well as, impact-parameter dependent
$\gamma\gamma$-luminosities. Excitations, like
the multiphonon giant dipole resonances, vector meson production and
multiple $e^+e^-$ pair production can be treated analytically in a
bosonic model, analogous to the emission of soft photons in QED.
\end{abstract}
\end{frontmatter}

\section{Introduction}
\label{sec:Introduction}

In ultraperipheral relativistic heavy ion collisions (UPC)
 very strong electromagnetic fields are present for a very short 
time \cite{beba88}.
The parameter which characterizes the strength of the interaction is 
the Coulomb parameter $\eta$:
\begin{equation}
\eta=\frac{Z_1 Z_2 e^2}{\hbar v} \approx Z_1 Z_2 \alpha
\end{equation}
where $Z_1$ and $Z_2$ are the charges of the nuclei and $v\approx c$
is the relative velocity of the ions. For heavy nuclei like Au-Au
(RHIC) or Pb-Pb (LHC) $\eta \gg 1$. One can treat these systems
semiclassically, with the nuclei following classical Rutherford
trajectories.  At high energies they are well approximated by straight
lines with impact parameter $b$.  In addition to elastic scattering,
which is due to the exchange of many photons, there are various kinds
of inelastic processes, like nuclear excitations, photon-nucleus
interactions leading to baryonic or mesonic excitations and
photon-photon processes. One can calculate the amplitudes for these
processes in the same way as the corresponding ones for real photons
\cite{BaurHTSK,Krauss}.

Probabilities for the excitation of the giant dipole resonance (GDR)
and $e^+e^-$-pair production are very large, of the order of one for
small impact parameter \cite{BaurHTSK}.  This means that these
processes will also occur simultaneously with other interesting
processes, where the probabilities are generally much smaller than
one.  The GDR is a strongly collective mode in nuclei.  The theory of
relativistic heavy ion excitation of one- and multiphonon GDR states
is given in \cite{babe86}.  There is also mutual excitation, where
each ion is excited to the (one phonon) GDR state.  This process has
been recently studied experimentally at RHIC \cite{sebastian}; it is
important for the luminosity determination and for triggering on
UPCs\cite{Baltz98,Adleretal}.  The fact that the double phonon giant
dipole resonance is excited strongly in a two-photon excitation
mechanism is of special interest for nuclear structure physics.  The
properties of these new collective modes could be explored with this
method \cite{HE94}.

The probability of electron-positron pair production calculated in
lowest order perturbation theory exceeds the unitarity limit of one
\cite{he+-,HenckenTB95b,Aste02}.  This means that multiple $e^+ e^-$
pair production is expected to be appreciable for heavy ions at RHIC
and LHC conditions.

The probability $P(b)$ of other inelastic processes is in general much
smaller than one. An important process is diffractive vector meson
production (especially $\rho^0$-production).  Probabilities for
$\rho^0$-production in close (grazing) collisions are about 0.01 or
0.03 for RHIC and the LHC, respectively.  The probability of the
multiple (double) electromagnetic production of $\rho^0$ mesons is of
the order of $10^{-3}$ to $10^{-4}$\cite{skjn99}.

One example of simultaneous processes is giant dipole resonance
excitation (followed by neutron decay) accompanied by
$\rho^0$-production.  This process was recently studied experimentally
at STAR\cite{Adleretal}.

This paper will present a general, unified treatment of multiphoton processes
in heavy ion collisions, focusing on the correlations which occur because all
of the processes share a common impact parameter, $\vec{b}$.  
Section~\ref{sec:theory} reviews the semiclassical treatment and then 
introduces multiphoton interactions.
Section~\ref{sec:examples} presents some examples of multiphoton processes 
and gives examples of the correlations, with RHIC and the LHC used for 
illustration. We give our conclusions in Section~\ref{sec:conclusions}. A 
preliminary account of part of the present work is given in 
Ref.~\cite{BaurErice02}.

\section{Theory}
\label{sec:theory}
 
We work in the rest frame of one of the nuclei (``target
system''). The other nucleus moves on a straight line with impact
parameter $b > R_1+R_2$, (where the nuclear radii are $R_1$ and $R_2$,
respectively).  The corresponding Lorentz-factor is
$\gamma=1/\sqrt{1-v^2/c^2}$.  It is related to the Lorentz-factor
$\gamma_{coll}$ in the collider frame by $\gamma=2\gamma_{coll}^2-1$.

Since the strength of the electromagnetic 
interaction decreases with increasing impact parameter
$b$, the excitation probability also decreases, in many 
important cases proportional to $1/b^2$.
Higher order effects
are proportional to the product of such probabilities, and are
therefore especially important for collisions with impact parameter
$b$ close to $b_{min}=R_1 + R_2$.

The collision time is given by $\tau_{coll}= b/\gamma$,
so the electromagnetic spectrum will include frequencies up to 
\begin{equation}
\omega_{max}\approx \frac{\gamma}{b_{min}}.  
\end{equation}

\subsection{Elastic Coulomb Collisions}
\label{sec:elastic}
\begin{figure}
\begin{center}
\resizebox{!}{1.5in}{\includegraphics{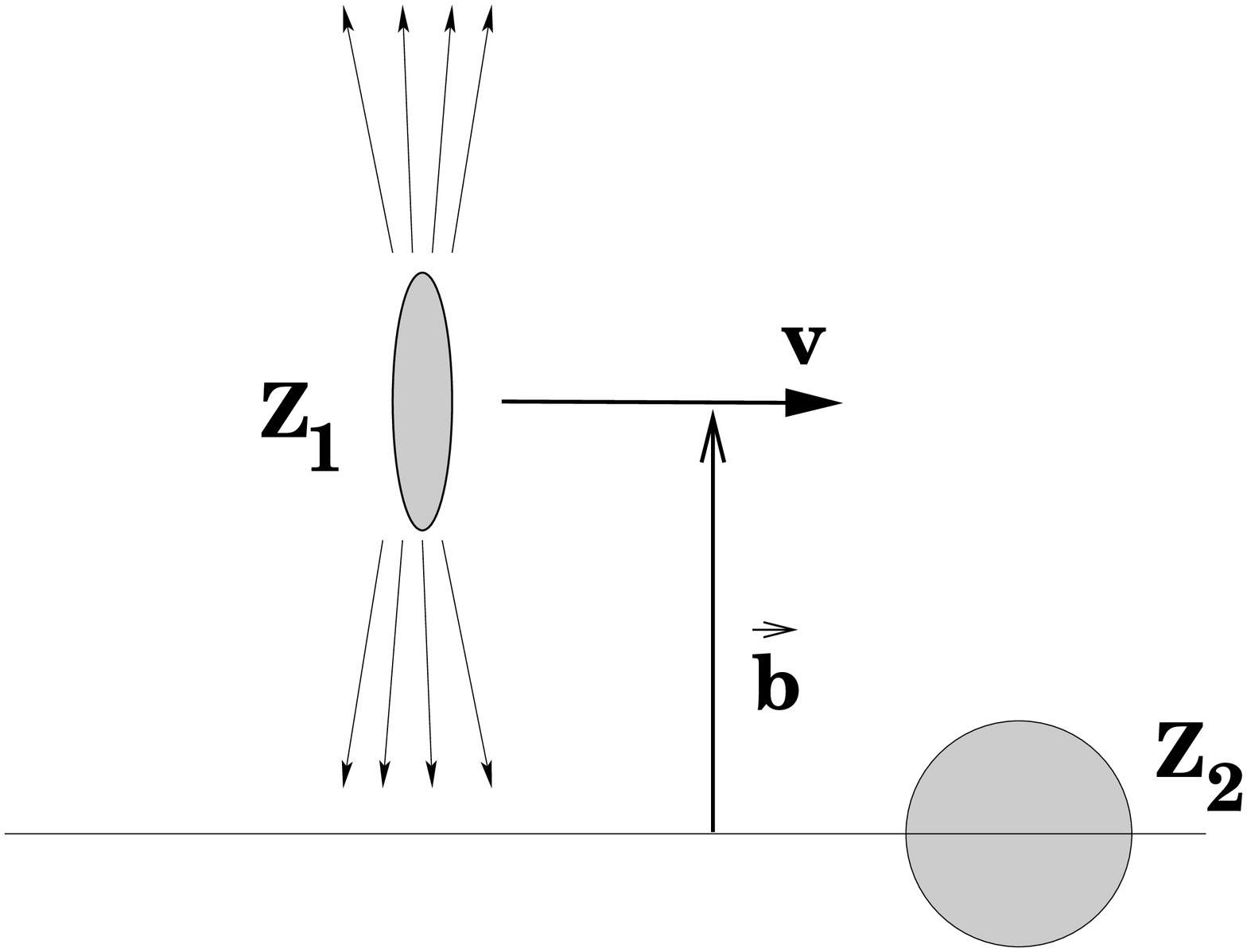}}
(a)\hfil
\resizebox{!}{1.5in}{\includegraphics{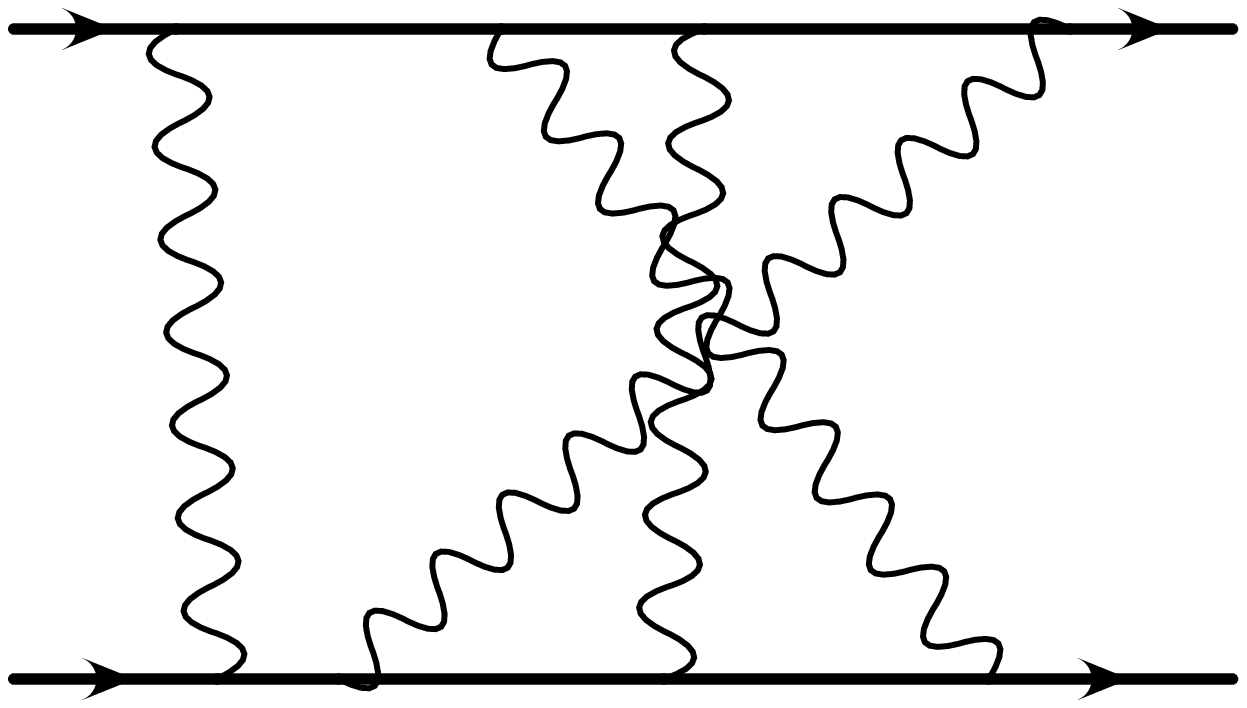}}
(b)
\end{center}
\caption{a) Classical picture of heavy ion scattering.
This picture was  also used by Fermi in his celebrated 
paper on the equivalent photon approximation
\protect\cite{Fermi}.
He considered the nonrelativistic case $v\ll c$. The relativistic
case is discussed in a 
very pedagogical manner in \protect\cite{JacksonED}. 
(b) For $\eta \gg1$ many photons are exchanged in 
elastic heavy ion scattering. A typical graph is shown.  
}
\label{fig:classglauber}
\end{figure}

A simple and rather accurate result 
is well known for classical  Coulomb scattering at small angles $\theta \ll 1$
\cite{JacksonED}. The
momentum transfer is
\begin{equation}
\Delta p=\frac{2Z_1Z_2e^2}{bv}=\frac{2\eta \hbar}{b}
\label{eq:classel}
\end{equation}
in the direction perpendicular to the motion. The momentum transfer
is related to the scattering angle $\theta$ by $\Delta p \approx p \theta$.
From this classical relation between the impact parameter and 
the scattering angle one can calculate the classical relativistic
Rutherford  scattering cross section as 
\begin{equation}
\frac{d\sigma}{d\Omega}=\left[\frac{2Z_1Z_2e^2}{p v}\right]^2 
\frac{1}{\theta^4},
\end{equation}
where $p=\gamma M_1 v$ and $M_1$ is the mass of nucleus $Z_1$,
see, e.g., p.~644 of \cite{JacksonED}.
This result can also be obtained in a field theoretical approach,
which we only sketch here.
Coulomb scattering is due to the exchange of photons. 
The sum over all the ladders diagrams, 
crossed and uncrossed, can be done in the high energy 
limit and for forward angles \cite{LS69,TT70,HI72}.
A typical diagram is shown in Fig.~\ref{fig:classglauber}(b).

One 
ingredient in this procedure is the linearization of the 
propagator denominator  
\begin{equation}
\frac{1}{(p+q)^2-m^2} \approx \frac{1}{2 p\cdot q}
\end{equation}
In order to sum up all ladder and crossed ladder graphs one uses the identity
\begin{equation}
\sum_{\sigma} \frac{1}{a_{\sigma(1)}(a_{\sigma(1)}+ a_{\sigma(2)})
\cdots (a_{\sigma(1)}+a_{\sigma(2)} + \cdots + a_{\sigma(n)})}
=\frac{1}{a_1 a_2 \cdots  a_n}
\end{equation}
where the sum is over all permutations $\sigma$ of the indices $1$ to $n$.

Summing over all orders the scattering amplitude is found similar to the
eikonal approximation in (non relativistic) potential scattering:
\begin{equation}
f_{el}(q) = \frac{p}{2\pi i} \int d^2b \exp(i \vec q \cdot \vec b) 
\left[ \exp(i \chi(b)) -1\right].
\label{eq:fel2}
\end{equation}
By integrating over the angle $\phi$
\begin{equation}
f_{el}(q) = i p \int_0^{\infty} db\; b J_0(qb) \left[1- \exp\left( i \chi(b)
\right)\right].
\end{equation}
In this  
approach there is a well defined momentum transfer
$q=\Delta p$ (we set $\hbar =1$), and
the impact parameter $b$ is not an observable quantity.
For the elastic Glauber phase $\chi(b)$ in the case of photon exchange, 
the analytic expression is well known, see e.g. \cite{Bartos02}:
\begin{equation}
\chi(b) =  -2\eta K_0(\lambda b),
\end{equation}
where $\lambda$ is a small photon mass needed to regularize the integral 
at large $b$.

For $\eta \gg 1$ the integrand oscillates very rapidly 
and large contributions to the integral are only 
obtained when the phase of this integrand is stationary.
One can introduce the saddle point (``stationary phase'') approximation
\cite{MorseF53}
\begin{equation}
\int e^{i \Phi(t)} dt \approx e^{i \Phi(t_0)} \sqrt{\frac{2\pi i}{\Phi''(t_0)}}
\label{eq:saddle}
\end{equation}
where $t_0$ is determined by  $\Phi'(t_0)=0$.
(One may have to sum over various stationary points $t_0$). 
In our case we have a two-dimensional integration over $\vec b$;
for a derivation of the semiclassical case using only a 
one-dimensional saddle point approximation, see \cite{BertulaniCG02}. 
Assuming that the momentum transfer $\vec q$ points in the $x$-direction, 
the phase in the integrand of Eq.~(\ref{eq:fel2}) is (for small $\lambda$)
\begin{equation}
\Phi(\vec b)=q_x b_x + 2\eta ln(\lambda b)
\label{eq:phib}
\end{equation}
The conditions
\begin{equation}
\frac{\partial\Phi}{\partial b_y} = 2 \eta \frac{b_y}{b^2} = 0,\quad
\frac{\partial\Phi}{\partial b_x} = q_x  + 2 \eta \frac{b_x}{b^2} = 0
\end{equation}
lead on the one hand to $b_y=0$, that is the dominant contribution
comes from the direction, where $\vec b$ is parallel to $\vec q$. 
It also leads to an extremum for $b_x=-b_0$ with
\begin{equation}
\label{eq:classical}
q=\Delta p=\frac{2\eta}{b_0},\qquad b_0=\frac{2\eta}{q}.
\end{equation}
The second derivatives at the saddle point are 
\begin{equation}
\frac{\partial^2 \Phi}{\partial b_x^2} = \frac{-2\eta}{b_0^2},
\frac{\partial^2 \Phi}{\partial b_x \partial b_y} = 0,
\frac{\partial^2 \Phi}{\partial b_y^2} = \frac{2\eta}{b_0^2}.
\label{eq:phipp}
\end{equation}
Therefore an approximation of the elastic scattering amplitude 
up to a phase is
\begin{equation}
|f_{el}(q)| = \frac{p b_0^2}{2\eta} = \frac{2 p \eta}{q^2}, 
\end{equation}
giving the same cross section as the classical calculation.

Eq.~(\ref{eq:classical})  gives the 
classical connection of the momentum transfer $q$ to the impact parameter
$b_0$, see above Eq.~(\ref{eq:classel}). 
The dominant contribution to the (pseudo-) Gaussian integral comes from 
values of $\vec b$ around the classical impact parameter $b_0$ in 
(opposite) direction of $\vec q$,
with a width (``uncertainty of the $b$-value'', see Eq.~(\ref{eq:phipp})) 
of $\frac{b_0}{\sqrt{\eta}}$.

The classical momentum transfer $\Delta p$ is built 
up from the perpendicular momenta of many exchanged photons.
The perpendicular momentum of the individual photon is restricted to
$q_\perp \lesssim \frac{1}{b}$, so 
$2\eta$ photons must be exchanged to reach the classical
$\Delta p$.  This contrasts with electron or proton
scattering, where $\eta =\frac{1}{137}\ll 1$ for $v \approx c$.
So, in the following, we use the semiclassical 
approximation to describe inelastic processes in the case $\eta \gg 1$.

For $b<R_1+R_2$ one has to consider strong interactions.  In many
cases, especially for heavy ions, the ``black disk approximation'' is
quite good. In this approximation $\chi$ is very large and imaginary
for $b < R_1+R_2$. In the case of strong Coulomb elastic scattering 
($\eta \gg 1$), including the strong
absorption of the ions for $b<R_1+R_2$, the diffraction pattern is
given by Fresnel diffraction (rather than a Fraunhofer one for $\eta
<1$). This is explained in detail in \cite{Frahn72}, see also
Sec.~5.3.5 of \cite{NoeWei}.  These grazing angles are extremely
small, about $5\times 10^{-5} $ (radians) for Au-Au collisions RHIC
and $2\times 10^{-6}$(radians) for Pb-Pb collisions at LHC (in the
collider frame); they are well below the beam emittance, so the
scattered ions will remain in the beam.
Although the previous discussion is not entirely new, we think that it is
useful to present the general ideas, which relate classical and quantal
(or field-theoretical) descriptions of scattering in strong Coulomb fields,
which may be a rather unfamiliar subject.

\subsection{Perturbation Theory for Inelastic Processes}
\label{subsec:inel}

In the following we use the semiclassical approximation and treat the
Coulomb field of the ion(s) as classical external fields.  One ion
moves on a straight line trajectory with a definite impact parameter,
while the other ion is at rest.  The semiclassical approximation can
be justified in the same way as in the elastic case.  The scattering
amplitude in the inelastic case is given in the eikonal approximation
by \cite{BertulaniCG02}
\begin{equation}
f_{fi}(\vec q)=\frac{p}{2\pi i}\int_0^\infty d^2b \exp(i \vec q \cdot \vec b)
\exp [i\chi(b)] a_{fi}(\vec b).
\end{equation}
The amplitude $a_{fi}$ can be expanded in the form
$a_{fi}(\vec b) = \sum_{\mu} a_{fi}^{\mu}(b) \exp(i\mu\phi)$
where $\mu$ denotes the angular momentum transfer along
the beam axis. 
After integration over the azimuthal angle $\phi$,
\begin{equation}
f_{fi}(\vec q)=-ip \int_0^\infty db \sum_{\mu} J_\mu(q b) i^\mu
\exp [i\chi(b)] a_{fi}^{\mu}(b).
\end{equation}
Using the saddle point approximation, the connection to
the semiclassical case can clearly be seen and $a_{fi}(\vec b)$ corresponds
to the semiclassical excitation amplitude.
In the semiclassical approximation, the scattering amplitude
is the product of the elastic  Coulomb scattering amplitude
and the semiclassical excitation amplitude $a_{fi}(b)$. 
We have 
\begin{equation}
f_{fi}(\vec q)=f_{el}(q) a_{fi}(- b_0 \hat q) = f_{el}(q) \sum_\mu (-1)^\mu
a_{fi}^\mu(b_0)
\label{eq:ffim}
\end{equation}
where  $\hat q$ is the unit vector pointing in the $\vec q$-direction
(see Eqs.~(\ref{eq:phib})--(\ref{eq:phipp})).
While we have not made numerical comparisons of the field theoretic
solution to the semiclassical approximation in the relativistic case,
such comparisons exist for the non-relativistic electromagnetic
excitation. In this case the electromagnetic cross sections can be 
calculated exactly. They depend essentially on the 
Coulomb parameter $\eta$ which approaches infinity in the 
semiclassical approximation. It is found (see, e.g., Fig.~(II.9)
of \cite{AlderBHM56}) that the asymptotic (semiclassical) 
limit is reached already for rather low $\eta$ values.
The semiclassical approximation works better than one may expect. 
In Au-Au or Pb-Pb collisions $\eta \approx 50$, so
the semiclassical approximation should be excellent.

The electric charge of the relativistic ion gives rise to
an electromagnetic potential, the Lienard-Wiechert
potential $A_{\mu}(\vec r,t)$ 
\begin{eqnarray}
A_0(\vec r,t) &=& \phi(\vec r,t)=
\frac{Z_p e \gamma}{\sqrt{(b-x)^2+y^2+\gamma^2(z-vt)^2}}\nonumber\\
\vec A(\vec r,t) &=& \frac{\vec v}{c} \phi(\vec r,t).
\end{eqnarray}
The interaction with the target is described in terms of an
electromagnetic current operator $\hat j_{\mu}$.  This leads to a 
time-dependent electromagnetic interaction of the form
\begin{equation}
V(t)=\int d^3r A_{\mu} (\vec r,t) \hat j^\mu (\vec r).  
\label{eq:vdefined}
\end{equation}
The dependence of $V(t)$ on the impact parameter is not shown 
explicitly.
The target current $\hat j_\mu$ describes a wealth of 
physics and is a quite complicated object in general. This current 
contains the current of the nucleus and the 
interaction leads to the excitation of nuclear states, with
the giant dipole resonance (GDR) as an important example. 
It also describes  photonuclear interactions like nucleon excitations,
meson production and photon-photon interactions, for example lepton
pair production.
The lepton ($e^+e^-$) states are those in the external Coulomb field
of the target, i.e., the Furry picture is used here.
Now one can expand the (time dependent) target state $\Phi(t)$ 
in terms of all possible states $\Phi_n$ which can be reached 
by the interaction $V(t)$. We write
\begin{equation}
\Phi(t)=\sum _n a_n(t)  \exp(-iE_n t) \Phi_n,
\end{equation} 
where we have introduced the time-dependent amplitudes $a_n(t)$ for the
state $\Phi_n$. The (time-independent)
states $\Phi_n$ consist of all states which can be connected 
to the target nucleus ground state by the interaction $V$
(i.e., due to the interaction with the (virtual) photons). 
This includes, for example, excited nuclear states, the nucleus 
in its ground state and lepton pairs, or mesons, or 
nuclear excited states along with lepton pairs, or mesons, etc.

One can set up coupled equations for the amplitudes $a_n(t)$ \cite{LandauLQED}.
They are
\begin{equation}
i \dot a_n=\sum _m \left<n\right|V(t)\left|m\right> \exp(i(E_n-E_m)t) a_m(t)
\label{eq:andot}
\end{equation}
The initial condition  is $a_m(t \rightarrow -\infty) =\delta_{m0}$.
The expression on the r.h.s. can be viewed as
the matrix element of the operator
\begin{equation}
\tilde V(t)=\exp( iH_0 t) V(t) \exp(-iH_0 t)
\label{eq:vtilde}
\end{equation}
(``interaction representation''). A formal solution of this equation 
can be written down using the time-ordering operator ${\mathbbm T}$:

\begin{equation}
a_n(t\rightarrow +\infty) = 
\left< n \right| {\mathbbm T} \exp\left(-i
\int_{-\infty}^{+\infty} dt \tilde V(t) \right) \left| 0 \right>.
\label{eq:antordered}
\end{equation}

We use perturbation theory to solve this equation.
Since the electromagnetic interaction $V$ is in general weak,
this is a good procedure. The first order amplitude is
\begin{equation}
a_n^{(1)}=-i \int_{-\infty}^{\infty} dt 
V_{n0}(t)\exp(i\omega_{n0}t)
\end{equation}
where 
$V_{nm}(t)=<n|V(t)|m>$
and $\omega_{nm}=E_n -E_m$.
This is the one-photon approximation. In second order
\begin{equation}
a_n^{(2)}= \frac{1}{i^2} \sum _m \int_{-\infty}^{\infty} dt
V_{nm} (t) \exp\left(i\omega_{nm}t\right) \int_{-\infty}^t dt'V_{m0}(t')
\exp\left(i\omega_{m0}t'\right),
\end{equation}
where the sum is over all intermediate states $m$. Only if 
the integrand is symmetric in $t$ and $t'$ one can extend the 
integral over $t'$ to infinity and factorize the amplitude. 
This is not possible here since the operator $\tilde V$ does not
commute in general for different times $t$ and $t'$.
(In the case of the excitation of a harmonic oscillator,
the commutator of these operators is a $c$-number and a full analytical 
solution can be found; this case will be treated below in 
Sec.~\ref{subsec:dgdr}.)

\begin{figure}
\begin{center}
\resizebox{!}{1.5in}{\includegraphics{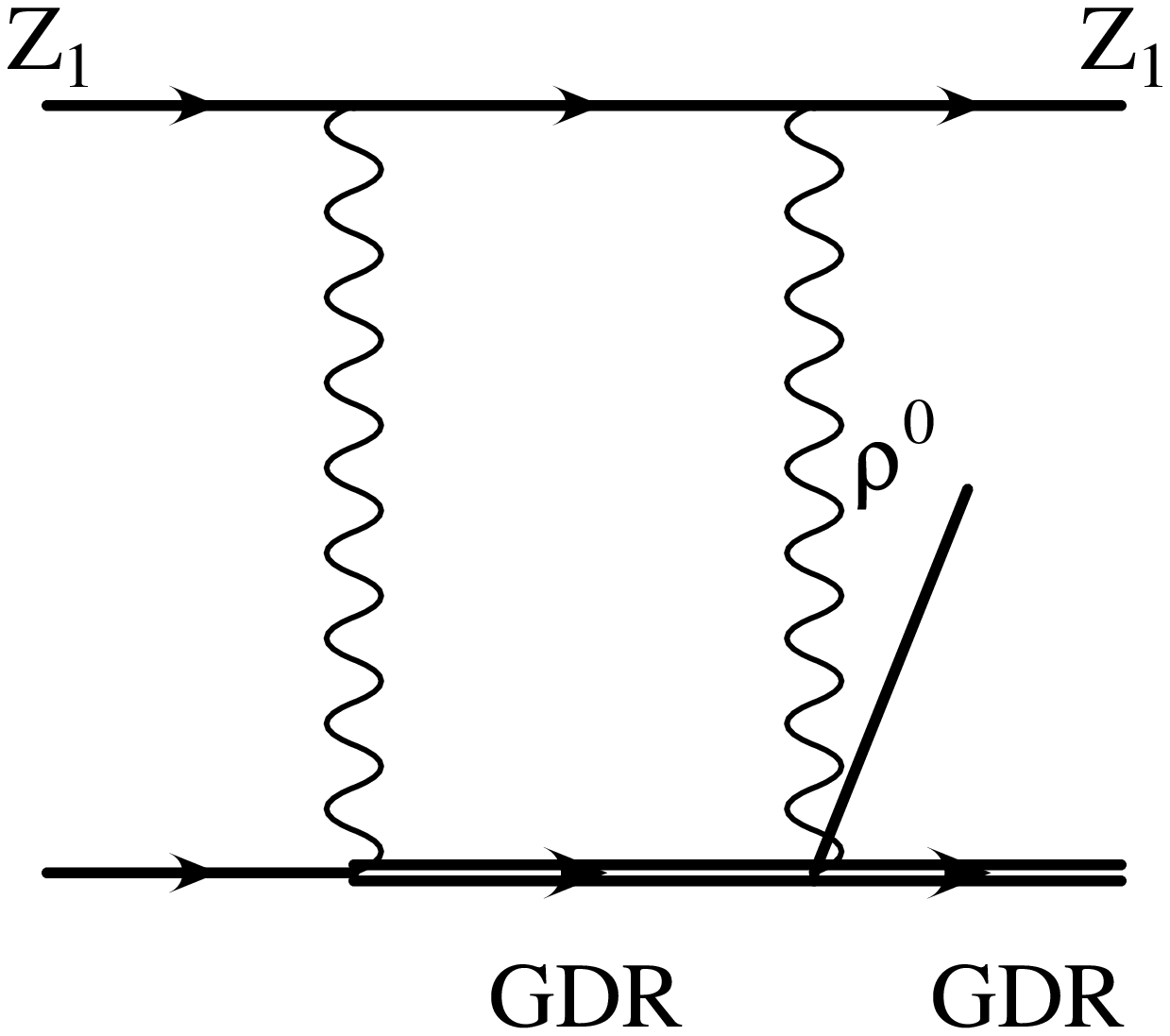}}
\hfil
\resizebox{!}{1.5in}{\includegraphics{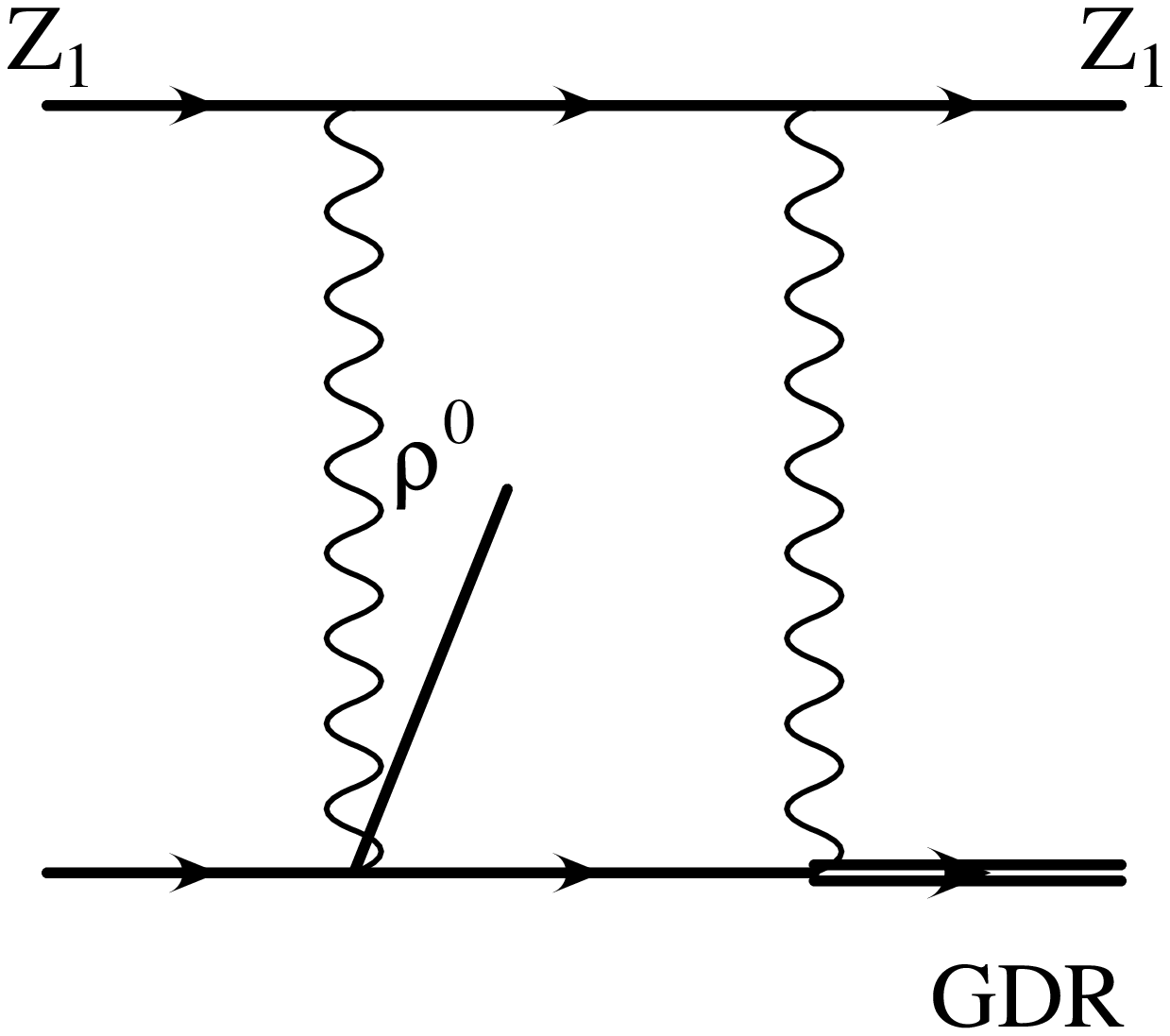}}
\end{center}
\caption{ Graphs contributing to the simultaneous 
excitation of states $\alpha$ and $\beta$. 
In this figure we take $\alpha=GDR$ and $\beta = \rho ^0$ as 
a typical example.
\label{fig:fig3}}
\end{figure}

In the case of the excitation of two independent modes, 
that is of two modes which do not mix with each other since their  
interaction is zero or negligible,
there is such a factorization: let us assume that one excites a state
$n=(\alpha, \beta)$ where $\alpha$ denotes, e.g., a nuclear excited state
and $\beta$ a vector meson in a state with
a given momentum and helicity produced coherently.
Two types of intermediate states
contribute to the sum over $m$, see Fig.~\ref{fig:fig3}:
$m=(0,\beta)$ and $m=(\alpha,0)$. 
By summing over them the integrand becomes symmetric in $t$ and $t'$. 
Formally this can then be written in the following way: 
The first path goes from the state $0=(0,0)$ to $(\alpha,0)$
and then to $(\alpha,\beta)$. The contribution of this path
to the second order  matrix element is 
\begin{eqnarray}
&&a_{(\alpha,\beta)}^{(2)}(1)= \nonumber\\
&&\frac{1}{i^2} \int_{-\infty}^{\infty} dt
V_{(\alpha,\beta)(\alpha,0)} (t) \exp\left(i\omega_{\beta 0}t\right) 
\int_{-\infty}^t dt' V_{(\alpha,0)(0,0)}(t')
\exp\left(i\omega_{\alpha 0}t'\right).
\label{eq:path1}
\end{eqnarray}
and for the second path 
\begin{eqnarray}
&&a_{(\alpha,\beta)}^{(2)}(2)= \nonumber\\
&&\frac{1}{i^2} \int_{-\infty}^{\infty} dt
V_{(\alpha,\beta)(0,\beta)} (t) \exp\left(i\omega_{\alpha 0}t\right) 
\int_{-\infty}^t dt' V_{(0,\beta)(0,0)}(t')
\exp\left(i\omega_{\beta 0}t'\right).
\label{eq:path2}
\end{eqnarray}
The independence of the two processes corresponds to the
assumption that the presence of the state $\alpha$ (or $\beta$) does
not influence the interaction matrix element for the production of
$\beta$ (or $\alpha$)
(and that a single photon cannot produce a single step transition
from $(0,0)$ to $(\alpha,\beta)$, see below):
$$
V_{(\alpha,\beta)(0,\beta)} = V_{(\alpha,0)(0,0)} = V_{\alpha 0}
$$
and
$$
V_{(\alpha,\beta)(\alpha,0)} = V_{(0,\beta)(0,0)} = V_{\beta 0}.
$$

In many important practical cases this independence should be very well
fulfilled. For mutual excitation the factorization is well understood,
the elastic form factor of the excited ion is almost
unchanged from the one of the ion in its ground state. Even
the slight change in form factor takes time, as will be discussed below
in connection with the validity of the sudden limit.

Similarly for photon-photon processes like
lepton-pair production together with nuclear excitation, 
the photon-photon process depends only
on the charge of the nucleus, with little sensitivity to the details of the
charge distribution. The nucleus is not influenced by the 
additional photon-photon process and vice versa.
For hadronic final states, where there may be strong interactions with
the nuclei, the photon-photon process in a
double equivalent photon approach will predominantly occur outside 
the two nuclei.

Adding the amplitudes of the two paths, Eqs.~(\ref{eq:path1}) 
and~(\ref{eq:path2}) one can see
that the integrand is symmetric in t and $t'$ and one obtains factorization.
In an obvious way this is generalized to the excitation of $N$
independent modes. For any function $f(t_1,t_2,\cdots,t_n)$ which is 
symmetric in all variables $t_i$ 
\begin{eqnarray}
&&\int_{-\infty}^t dt_1 \cdots \int_{-\infty}^t dt_n f(t_1,\cdots t_n)=
\nonumber\\ &&
n! \int_{-\infty}^t dt_1 \int _{-\infty}^{t_1} dt_2 \cdots
\int_{-\infty}^{t_{n-1}} dt_n f(t_1,t_2, \cdots t_n).
\end{eqnarray}

One can use this formula to factorize the amplitude
into a product of first order amplitudes. The factor 
of $\frac{1}{n!}$ is compensated by the $n!$ different ways to
order the excitation of the independent modes $\alpha,
\beta, \gamma ,\cdots$.

Let us give here also an example of two modes, which are coupled
(i.e. no longer independent): 
the excitation of vibrational and rotational modes of a nucleus
\cite{Baur89}. These modes mix due to rotation-vibration coupling. 
This gives rise to nuclear eigenstates of a mixed
character. In the case of $\rho^0$ production and 
GDR excitation, for example, such a coupling is 
expected to be so small that it can safely be neglected.

\begin{figure}
\begin{center}
\resizebox{!}{1.5in}{\includegraphics{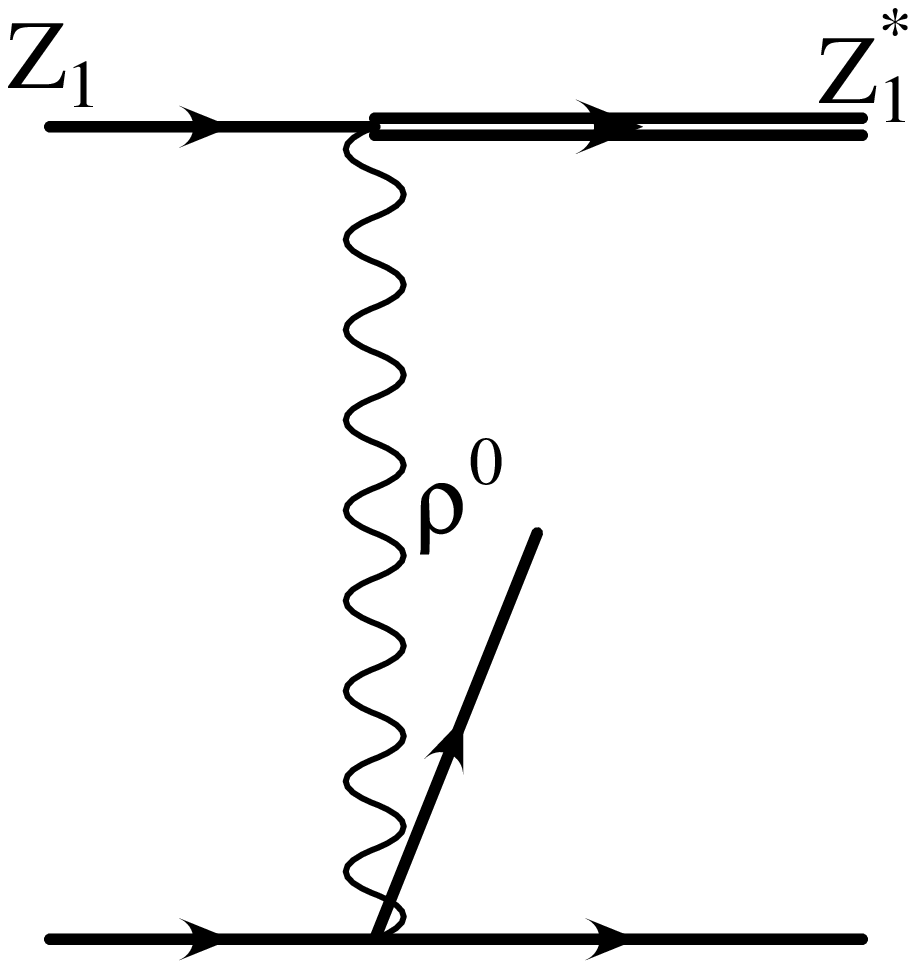}}
(a)~~~~~~~~~~~~~
\resizebox{!}{1.5in}{\includegraphics{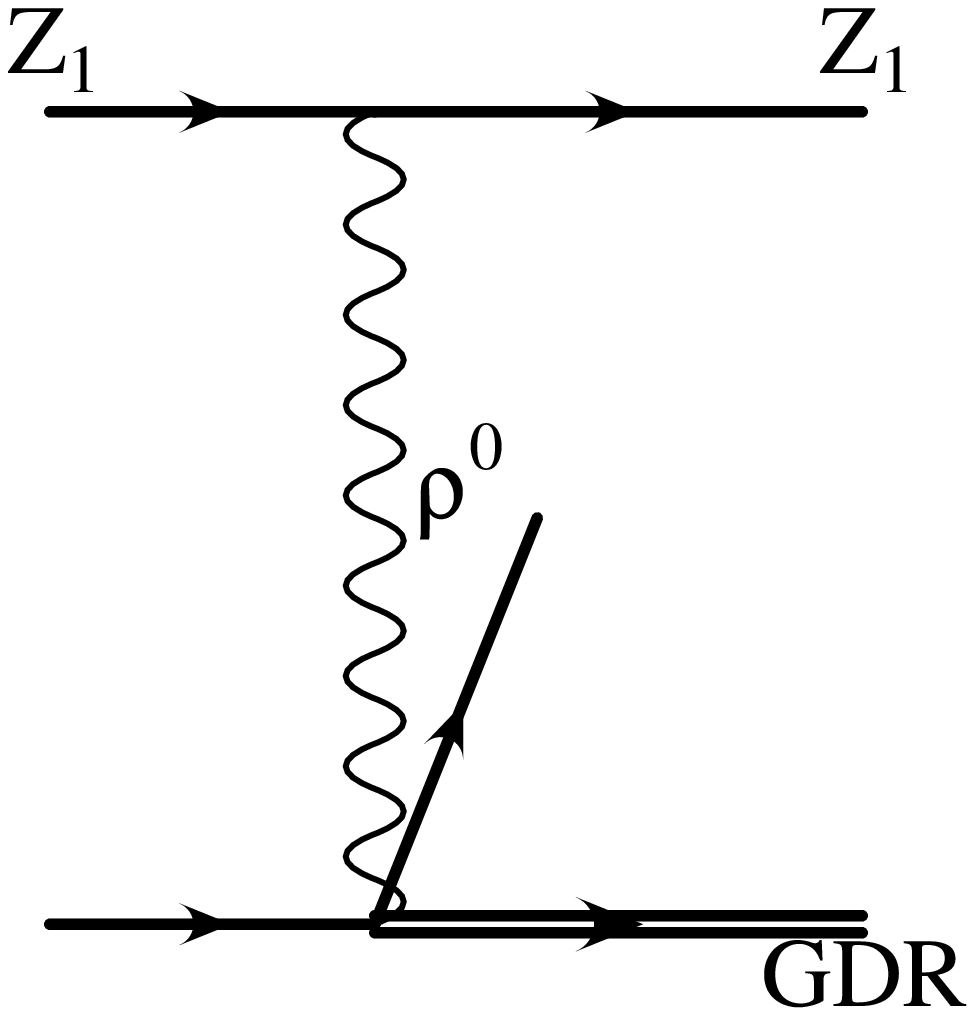}}
(b)
\end{center}
\caption{Examples of graphs which spoil the factorization of the
different processes, compare also with the figure above. As discussed
in the text their contribution are expected to be small.
\label{fig:nonfac}}
\end{figure}

The factorization is also not fulfilled in single photon exchange processes,
which are depicted in Fig.~\ref{fig:nonfac}.
One such process is photon exchange with inelastic vertices on both sides
(Fig.~\ref{fig:nonfac}(a)). 
This process was studied in \cite{HenckenTB95,HenckenTB96} within an
inelastic EPA approach and was found
to be only of the order of 1\% as compared to the elastic process.
Another example is the GDR excitation together with the meson 
production, see Fig.~\ref{fig:nonfac}(b). This is a special contribution to 
the incoherent meson production process. Again this process is expected 
to be small: the vector meson is mainly produced through a 
diffractive (absorptive) process,
which is very ineffective for a GDR excitation. The magnitude
of such a process could be estimated by looking, e.g., at the GDR excitation
in inelastic meson scattering at high energies. Unfortunately we are unaware
of any study that has been done at these high energies.

Asymmetric systems, such as $d$-A are also of interest, as shown by
the recent $d$-Au run at RHIC, where UPCs were studied \cite{falk}.  In
these collisions, non-factorizable processes could come into play.
The Coulomb parameter is $\eta=0.6$, less than one and the
semiclassical approximation (used, i.e., in equations like
Eqs.~(\ref{eq:classel}), (\ref{eq:saddle}) or~(\ref{eq:ffim})) does
not apply. However, the total cross-sections (i.e. integrated over $b$
and $q$, respectively) are equal in the semiclassical and Glauber (or
plane wave) case \cite{beba88,ba91}.  In this run, the $\rho^0$+
nuclear breakup channel was studied.  Due to the $Z^2$ dependence the
photon(s) will come predominantly from the gold nucleus. The $\rho^0$
+ breakup channel gets the main contribution from incoherent rho
production, i.e. it is a one photon process of the form of
Fig.~\ref{fig:nonfac}(b).  There is also a two-photon contribution
from coherent $\rho^0$-production accompanied by an electromagnetic
deuteron breakup \cite{skleinrvogt}.  This mechanism is suppressed by
a factor of $P_{breakup}(b)$ relative to the incoherent one-photon
mechanism, which is of the order of 1\% in the relevant impact
parameter range.  Due to the selection mechanism in the experiment the
incoherent rho meson production mechanism is disfavored. Therefore
overall the balance between the two reaction mechanisms should be
studied in more detail.

\subsection{
Multiphoton Processes in the Sudden or Glauber Approximation}
\label{sec:multi}

The solution of the coupled equations Eq.~(\ref{eq:andot}) is  greatly 
facilitated if the sudden approximation can be applied,
see, e.g., \cite{ba91}. In this case the collision
time $t_{coll}=b/\gamma$ is much smaller than the excitation
time $1/\omega$. This condition is fulfilled in many interesting
cases and we assume now that the sudden approximation
can be used. This ``frozen nucleus''-approximation 
is also used in  Glauber theory. The relation between the 
semiclassical approach  and the Glauber 
(or eikonal) approximation is explained in \cite{ba91},
for the non-relativistic, as well as, for the relativistic case.
The excitation amplitude is 
\begin{equation}
a_n(t \rightarrow \infty) =<n| \exp(iR)|0>
\label{eq:ansudden}
\end{equation}
where 
\begin{equation}
R=-\int_{-\infty}^{+\infty} dt V(t)
= -\int_{-\infty}^{+\infty} dt A_\mu(t) \hat j^\mu.
\end{equation}
The operator $R$ is a direct sum of operators in
the space of nuclear states, the space of the 
nucleus-vector meson system, the nucleus-$e^+e^-$ system, etc.. 
So,
\begin{equation}
R = \sum_i R_i  
\end{equation}
where $i$ denotes the different final states like $e^+e^-$ pairs, 
$\rho$-mesons or a GDR excited ion. One has
\begin{equation}
\exp(i R) = \prod_i \exp(i R_i).
\label{eq:expirprod}
\end{equation}
as the different $R_i$ commute with each other.
One may expand the exponential in Eq.~(\ref{eq:ansudden}).
Terms linear in $R$ describe, e.g., the excitation of nuclear states, like the 
collective giant dipole resonance (GDR), 
vector meson production or $e^+e^-$pair production.
Terms quadratic in $R$ give, e.g., contributions to double-phonon
GDR excitation, double vector meson production,
double $e^+e^-$ pair production. The quadratic terms also describe, 
e.g., vector meson production and GDR excitation
in a single collision.
A contribution to the second order amplitude $a^{(2)}$ is, e.g., 
\begin{equation}
a^{(2)}=-<GDR|R|0><\rho^0|R|0>
\label{eq:a2}
\end{equation}
where $|0>$ denotes the ground state of the nucleus.
In the general case the $R_i$ commute
and using Eq.~(\ref{eq:expirprod}) one gets
\begin{equation}
a=<GDR|\exp(i R_{GDR})|0><\rho^0|\exp(i R_{\rho0})|0>,
\end{equation}
which has the second order term as a special limit.

For three independent processes, say GDR excitation, vector meson- and
$e^+e^-$ pair production, the six different time orderings compensate
for the $1/3!$ factor in the expansion of Eq.~(\ref{eq:ansudden}).  In
this formalism, processes are independent and the elementary
amplitudes factorize, as one would have intuitively expected.  This
property is used also in the experimental analysis of vector meson
production with simultaneous GDR excitation, where the neutrons from
the GDR decay serve as a trigger on UPC \cite{Adleretal}.  The ion
motion is not disturbed by the excitation process. The reason is that
the kinetic energy of the ion is much larger than the excitation
energy.  Due to coherence, the quantity $x=\frac{\omega}{E}
=\frac{1}{RM_N A}$ is very small. One has $x< 4 \times 10^{-3},3
\times 10^{-4}$ and $1.4 \times 10^{-4}$ for oxygen, tin and lead ions
respectively.

The GDR excitation immediately affects the nucleon momenta, but the nucleons 
will take a finite time to adjust their positions. With an excitation energy 
of about 20~MeV, the oscillation time of the GDR can be estimated classically
to be about 10~fm$/c$. Due to the time dilatation, this corresponds to 
a time in the collider frame of about 1000~fm$/c$
at RHIC and even longer at the LHC. This time is long compared
to the time required to, for example, produce a $\rho^0$.

From the general discussion above of higher order theory we have seen that
this is not a necessary assumption to achieve factorization.

\subsection{
Excitation of a Harmonic Oscillator (Bosonic Modes)}
\label{subsec:dgdr}

An especially simple and important case
is the excitation of a harmonic oscillator.
It also has the virtue that it can be studied analytically.
As some excitation processes in
heavy ion collisions can be regarded as quasibosonic processes, such a model 
can serve as a good approximation. Some examples are discussed below.
Although well known (see, e.g., \cite{AlW75,Merzbacher}), we briefly present
the main results of the excitation of a harmonic oscillator by an external
force, this serves also to establish the notation.

In terms of the creation and destruction operators 
$a^{\dagger}$ and $a$ the Hamiltonian of the system is 
\begin{equation}
H=\omega\left(a^{\dagger}a+\frac{1}{2}\right)
\label{eq:hamil}
\end{equation}
where $\omega$ is the energy of the oscillator. The boson commutation
rules are $[a,a^{\dagger}]=1$ and $[a,a]=[a^{\dagger},a^{\dagger}]=0$.
Only one mode is shown explicitly, in general one has to 
sum (integrate) over all the possible modes.

One assumes that the operator $V$ (see Eq.~(\ref{eq:vdefined}) is linear 
in the creation and destruction operators: 
\begin{equation}
V(t)=f(t) a + f^*(t) a^{\dagger}.
\end{equation}
In this case one can calculate $\tilde V$ explicitly using the boson 
commutation rules given above and the expansion
\begin{equation}
\tilde V(t)= V(t)+ it [H_0,V(t)]+\frac{(it)^2}{2!}[H_0,[H_0,V(t)]]+...
\end{equation}
One finds
\begin{equation}
\tilde V(t)=f(t)e^{-i\omega t} a  + f^*(t) e^{i\omega t} a^{\dagger}
\end{equation}

Now one can convince oneself that the commutator of
$\tilde V$ at different times $t$ and $t'$ is a pure $c$-number. In this case  
one can disregard the time ordering operator in Eq.~(\ref{eq:antordered})
and obtain an exact analytical answer,
up to an unimportant overall phase factor (see, e.g., \cite{AlW75}).

The excitation operator is the integral of the interaction over time 
\begin{equation}
\int_{-\infty}^{+\infty}dt \tilde V =- (u\ a + u^*\ a^{\dagger})
\label{eq:r}
\end{equation}
where $u$ is the  $c$-number 
\begin{equation}
u=\int_{-\infty}^{\infty}dt f(t)\exp(- i\omega t).
\end{equation}
This leads to the excitation of a so-called coherent state, see 
\cite{glauber,Klauder}.
For the excitation of multiphonon states this is 
explicitly shown in \cite{bebapadova}. One has
\begin{equation}
a_n=<n|e^{-i(u*a^{\dagger}+ua)}|0>=
\frac{(-iu^*)^n}{\sqrt{n!}}e^{-\frac{1}{2}u u^*},
\label{eq:coherent}
\end{equation}
where the operator identity
$e^{A+B}=e^A e^B e^{-\frac{1}{2}[A,B]}$ was used, which 
is valid for two operators $A$ ($-iu^*a^{\dagger}$)  and 
$B$ ($-iua$) for which the commutator is a $c$-number.
See also the discussion of the forced linear harmonic
oscillator in \cite{Merzbacher}.

Let us look at a few cases where this model can be applied:
Electromagnetic excitation of nuclear states, especially the
collective giant multipole resonances was discussed
recently by Bertulani \cite{carlos}. 
The possibility to excite multiphonon GDR states is studied in 
\cite{babeerice}. 
Multiphonon GDR  also play a role in the electromagnetic excitation
of the ions in relativistic heavy ion collisions, for a recent
 reference see \cite{Pshenichnov98,Pshenichnov01}.
The parameter which describes
the probability $P_{GDR}(b)$ of GDR excitation is (see \cite{babe86})
\begin{equation}
P(b)=\frac{S}{b^2}
\label{eq:phi}
\end{equation}
with a simple parameterization for $S$
for the relevant high beam energies \cite{beba88} 
\begin{equation}
S = \frac{2\alpha^2Z_1^2N_2Z_2}{A_2m_N\omega} = 
5.45 \times 10^{-5} Z_1^2 N_2 Z_2 A_2^{-2/3} \mbox{fm}^2.
\label{eq:Sgdr}
\end{equation}
where $m_N$ denotes the nucleon mass, and the neutron-,
proton-, and mass-number of the excited nucleus are 
$N_2,Z_2,$ and $A_2$ respectively.
The excitation probability is inversely proportional 
to the energy $\omega$ $(\approx 80\mbox{MeV}A^{-1/3})$ of the GDR state. 
As expected, the soft modes are more easily excited.
Eq.~(\ref{eq:phi}) is based on the assumption that
the classical dipole sum rule (Thomas-Reiche-Kuhn sum rule)
is exhausted to 100\%. For the excitation of an $N$-phonon state,
a Poisson distribution is obtained. For heavy systems $P(b)\approx 
\frac{1}{2}$ for close collisions ($b \gtrsim R_1+R_2$).

Double $\rho^0$ production was studied in \cite{skjn99}. In addition
to the label $m$ for the magnetic substates of the GDR, there is a
continuous label (the momenta).  The probability to produce a
$\rho^0$-meson in a close collision is of the order 1-3\% for the
heavy systems.  It will be interesting to study these events
experimentally to see how the $\rho-\rho$ interaction affects the
simple harmonic oscillator description, especially for the production
of $\rho^0$'s with close enough momenta.

Multiple $e^+e^-$-pairs
can be produced in relativistic heavy ion collisions \cite{pra}.
This work used the  sudden (or Glauber) approximation and a
quasiboson approximation for $e^+e^-$ pairs.
Thus the Hamiltonian has the form of Eq.~(\ref{eq:hamil}), where a sum over 
the quantum numbers of the lepton pair has to be included.
Using a QED calculation (including Coulomb corrections in the 
Bethe-Maximon approach) for one pair production as an 
input, a Poisson distribution is obtained for multiple 
pair production. This is quite natural, since this problem 
reduces to the excitation of a harmonic oscillator
(the modes are labelled by the spins and 
momenta of the $e^+e^-$ pairs).
More detailed calculations have verified that the Poisson
distribution holds quite well for multiple pair production,
although slight deviations are expected \cite{HenckenTB95b,BaurHTSK}.
A characteristic dimensionless parameter for this problem is 
$\Xi= [(Z_1 Z_2 \alpha^2)^2]/(m_e b)^2$, where $m_e$ is 
the electron mass and $b>1/m_e$. Here, $\log{\gamma}$ should be large.

With heavy systems like Au-Au or Pb-Pb, when
$b \gtrsim 1/m_e$,  $\Xi\approx 1$. 
The impact parameter dependence of $e^+e^-$ pair production has been
studied numerically in lowest order QED\cite{he+-}. Only recently an
approximate analytical formula for the total pair production
probability in lowest order $P^{(1)}$ was found. 
When $1/m_e<b<\gamma/m_e$ 
\cite{lms}
\begin{equation}
P^{(1)}= \frac{28}{9\pi^2}\Xi(2\ln\gamma^2-3\ln(m_e b))\ln(m_e b)
\label{eq:p1}
\end{equation}    
The $N$-pair production probability decreases strongly with 
increasing impact parameter $b$ 
(approximately like $\sim b^{-2N}$, for $b>1/m_e$).
 Therefore the probabilities $P^{(1)}(b)$
should be known accurately for an impact parameter
range of $0<b< \mbox{several}\ 1/m_e$. 
Since the multiple pair production happens in a single collision,
with a given impact parameter, there are correlations
in the momenta of the outgoing pairs. This will be 
briefly discussed in Sec.~\ref{sec:examples} below.

$e^+e^-$ pair production is of great practical interest. It could be
useful as a trigger for UPC collisions at the LHC \cite{HenckenKS02a},
but also constitutes a background for the most central parts of the
trigger system of ALICE.  \cite{HenckenKS02b}.  
Electron or muon pair production might also be a way to measure the
$\gamma\gamma$-luminosity \cite{ShamovT02}.

\begin{figure}
\begin{center}
\resizebox{!}{2.5in}{\includegraphics{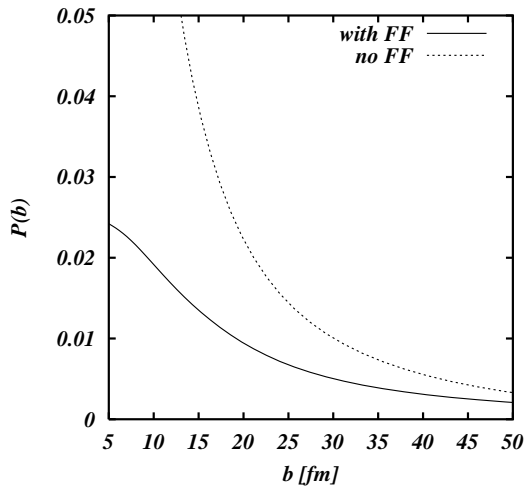}}
\end{center}
\caption{The probability of muon pair production is shown as a function of
impact parameter for Pb-Pb collisions at the LHC ($\gamma=3400$). Results 
for a monopole form factor $F(Q)=1/(1+Q^2/\Lambda^2)$ with $\Lambda=83$~MeV
and without a form factor ($F(Q)=1$) are given.
}
\label{fig:muon}
\end{figure}

Muon pair production is also of interest: in close collisions
the pair production probability is appreciable, although significantly 
smaller than one. This can be seen from Eq.~(\ref{eq:p1}).
The muon Compton wavelength $1/m_{\mu}$ is 1.86~fm, much 
smaller than the nuclear radius. For $b\approx 2R_A$
form factor effects also need to be considered.
Total cross sections are smaller by about a factor of $(m_e/m_{\mu})^2$
$\approx 2\times 10^{-5}$ (which is still large). 
The results of a lowest order external field QED calculation are shown in
Fig.~\ref{fig:muon}. Calculations with and without a monopole form factor
are shown together with the approximate expression of Eq.~(\ref{eq:dLggbW}) 
below. Details of the calculation are given in \cite{HenckenTB95b}.

\section{Some Examples of Multi-Photon Processes}
\label{sec:examples}
The factorization of the amplitudes has some significant,
experimentally accessible implications.  Although the amplitudes
are independent, the processes all share a common impact parameter
$\vec b$.  This leads to some significant correlations, due to both
the magnitude and direction of $\vec b$.  In this section, we consider
the effect of multiple interactions on the photon spectrum and impact
parameter distribution, and on the angular distributions of the final
states.

\subsection{Impact parameter dependence of the equivalent photon spectrum}
The photon spectrum at impact parameter $b>R_A$ is 
\begin{equation}
N(\omega,b) = {Z^2\alpha\over\pi^2} ({\omega\over\gamma})^2
\bigg(K_1^2(x)+{K_0^2(x)\over\gamma^2}\bigg)
\end{equation}
where $x=\omega b/\gamma$.  As long as $x \ll 1$, the photon spectrum
scales as $dN/d\omega \approx 1/b^2$.  The mean impact parameter
$\overline b$ is
\begin{equation}
\overline b = {\int d^2b\  b P(b) \over 
\int d^2b P(b)}
\end{equation}
where the probability $P(b)$ of a specific reaction occurring at impact
parameter $b$ is 
\begin{equation}
P(b) =  N(\omega,b) \sigma_{\gamma A}.
\end{equation}
For a reaction involving a single photon with energy $\omega$, we
can integrate over the region $b_{min} = 2R_A< |b|<b_{max} =
\gamma/\omega$. Many processes can occur over a range of
photon energies; in this case, an integral over $\omega$ is needed.
However, the details of the interaction do not matter, and
\begin{equation}
\overline b = {b_{max}-b_{min}\over \ln(b_{max}/b_{min})}.
\end{equation}
For single photon processes the average impact parameter $\overline b$
is large when $b_{max}$ is large and insensitive to the exact value of
$b_{min}$.  

However, for two or more photon interactions, such as mutual Coulomb 
excitation of two nuclei, $P(b) \approx (d^2N_\gamma/d^2b)^n 
\approx 1/b^{2n}$, with $n$ the number of photon interactions and the
latter approximation requiring that the $x$ for the two photons both
be less than one.  Then, as long as $b_{max}\gg b_{min}$,
\begin{equation}
\overline b_n \approx b_{min} \frac{2n-2}{2n-3}.
\end{equation}
$b_{max}$ drops out, and the median impact parameter only depends on
$b_{min}$, independent of the collision energy and other details of
the interaction. Of course, $b_{max}$ depends on the photon energy
and we assume that for all photon processes $b_{max}\gg b_{min}$ should be
fulfilled in the relevant $\omega$ range.
For heavy nuclei like Au $\overline b_2 \approx 2 b_{min} = 4 R_A = 26$~fm.

For three-photon interactions, such as vector meson production accompanied
by mutual Coulomb excitation of both nuclei, $n=3$ and 
$\overline b_3 \approx {4b_{min}/3} = {8R_A\over 3}$.
For heavy nuclei, $\overline b_3 \approx 18$ fm.  Of course, as
$\overline b_n$ drops with increasing $n$,
the cross sections for UPCs accompanied by
hadronic interactions become more sensitive to the tails of the
nuclear density profile --- including these tails increases $b_{min}$
above the canonical $2R_A$.

The $\overline b$ are comparable to the median impact parameters
calculated in \cite{BaltzKN02}; these detailed calculations found
that the median impact parameter is almost independent of the
specifics of the reaction.  This impact parameter distribution is
important for studying interference between vector meson production on
the two nuclei; the smaller $\overline b$, the easier it is to observe
the interference \cite{spenprl}.

The reduction in $\overline b$
affects the photon spectrum, since the maximum photon energy scales as
$1/b$.  One way to quantify this is to consider a two-photon reaction:
excitation of one nucleus to a GDR, while the second photon produces a
$\rho$ meson.  The GDR tagging has a strong effect on the
$\rho^0$-production: it selects small impact parameters, where the
equivalent photon spectra are harder.  One can calculate the equivalent
photon spectrum in the case of GDR tagging.  For close collisions,
which matter most, the probability of GDR excitation 
is given in Eqs.~(\ref{eq:phi}) and~(\ref{eq:Sgdr}).
The equivalent photon spectrum is then
\begin{equation}
n_{GDR}(\omega)=S \frac{ 2 Z_1^2\alpha}{\pi}(\frac{\omega}{\gamma v})^2
\int_{x_{min}}^\infty \frac{dx}{x}\left(K_1(x)^2+
\frac{K_0(x)^2}{\gamma ^2}\right).
\end{equation} 
This expression can be evaluated numerically.
Due to the high values of the Lorentz boost $\gamma$
at the colliders, $K_0(x)^2/\gamma^2$ is very small and can safely 
be neglected. A simple approximation 
\cite{BetheJ68} 
is found by setting $x^2 K_1(x)^2 =1$
for $x<1$ and zero otherwise; the integral over $x$ is then
$\frac{1}{2x_{min}^2}-\frac{1}{2}$
where $x_{min}=\frac{\omega R_{min}}{\gamma v}<1$. Then
\begin{equation}
n^{BJ}_{GDR}(\omega) \approx S 
\frac{Z_1^2\alpha}{\pi}(\frac{\omega}{\gamma v})^2
\left(\frac{1}{x_{min}^2}-1 \right).
\label{eq:ngdrbj}
\end{equation} 
The photon spectrum scales very roughly as
$n(\omega)\approx \omega^0$, in contrast to the
$n(\omega)\approx \omega^{-1}$ for untagged photons.

Using Eq.~(11.3.30) of Ref. \cite{AbramowitzS64} one finds a better
analytic expression
\begin{equation}
n_{GDR}(\omega) \approx \frac{S}{R_{min}^2}
 \frac{Z_1^2\alpha}{\pi}
 x_{min}^2 
(K_1(x_{min})^2 - K_0(x_{min})^2).
\label{eq:ngdras}
\end{equation} 

\begin{figure}
\begin{center}
\resizebox{!}{2.5in}{\includegraphics{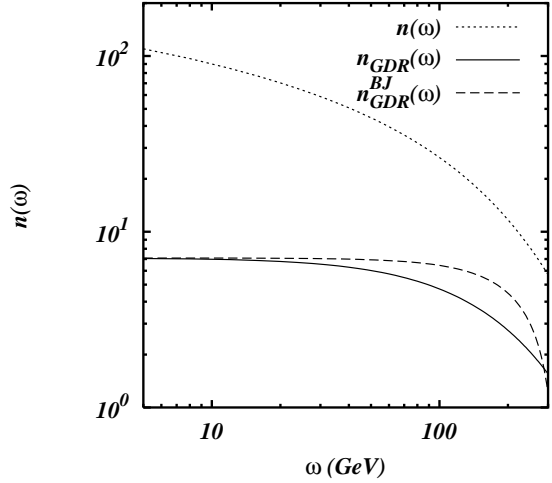}}
~~~
\end{center}
\caption{ The equivalent photon spectrum $n(\omega)$ (without tagging)
is compared to the one for the excitation of a GDR resonance in one of
the ions for Au-Au ($Z=79$, $A=197$) collisions at RHIC
($\gamma_{coll}=108$). The full expression is compared to the two
approximations given in Eqs.~(\ref{eq:ngdrbj}) and~(\ref{eq:ngdras})}
\label{fig:ngdr}
\end{figure}

Figure~\ref{fig:ngdr} shows
the EPA spectrum $n(\omega)$ (without tag)
and the spectrum $n_{GDR}(\omega)$ 
in  the  approximations of Eqs.~(\ref{eq:ngdrbj}) and~(\ref{eq:ngdras}) 
respectively. 

Another example of multiphoton processes is GDR tagging of
$\gamma\gamma$-processes: if the mass of the system produced in the
$\gamma \gamma$ interaction is low, there is a considerable influence
of the tagging, if the mass is high, the influence is less.
The $b$-dependence of (untagged) photon-photon-processes
can be found from the folding of $b$-dependent equivalent photon
spectra, see e.g. Eq.~(49) of Ref.~\cite{BaurHTSK}. A simple approximate
formula was developed recently for the special case of $e^+e^-$ pair
production.  The result is given in Eqs.~(0.23), (0.16) and~(0.17) of
\cite{lms}.

We consider here the more general case where
the invariant mass of the $\gamma \gamma$-system is given by $W$
(rather than $2m_e$, as in Ref. \cite{lms}).
The lower cutoff in the impact parameter $b_{min}$ is $2R_A$, 
(rather than the Compton wavelength $\frac{1}{m}$ of the electron).
We introduce the parameters $u=\frac{b}{R}$ and 
$x=\frac{2\gamma}{WR}$ . We reconsider Eq.~(0.24) of \cite{lms},
with different integration limits: $b_1$ from $R$ to $b$ and 
$\omega_1$ from $W^2 R/4\gamma$ to $\gamma/R$.
Generalizing Eqs.~(0.16) and~(0.17) of \cite{lms}, 
the $b$-dependent $\gamma\gamma$-luminosity is
\begin{equation}
\frac{d^3L_{\gamma\gamma}}{d^2b dW}= \frac{2\pi}{W R^2}
\left(\frac{Z^2\alpha}{\pi^2 u}\right)^2 X.
\label{eq:dLggbW}
\end{equation}
The probability $P_{AA}(b)$ of a $\gamma\gamma$-process in the 
AA collision is given in terms of 
this $\gamma\gamma$-luminosity by $P_{AA}(b)= 2 \pi b
\int \frac{dW}{W} \frac{d^3L_{\gamma \gamma}}{d^2bdW} \sigma_{\gamma \gamma}
$. In this expression $\sigma_{\gamma \gamma}$ is the elementary 
$\gamma \gamma$ cross-section and
\begin{equation} 
X= \left[\ln(x^2/u)\right]^2  \qquad\mbox{for $x<u<x^2$}
\end{equation}
(for ``large ''impact parameters) . For ``small'' impact parameters
\begin{equation}
X= 2\ln(u)(2\ln(x/u)+1/2\ln(u))   \qquad\mbox{for $2<u<x$}.
\end{equation}
This result is obtained by a generalization of Eq.~(0.24) of \cite{lms}.

\begin{figure}
\begin{center}
\resizebox{!}{2.5in}{\includegraphics{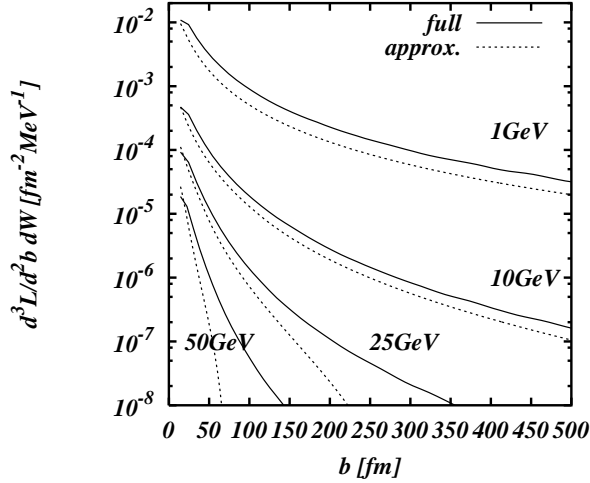}}
\end{center}
\caption{
The photon-photon luminosity $d^3L/d^2bdW$ is shown for different values of
$W$ as a function of the impact parameter $b$. The results of a full 
calculation for lead beams at the LHC are compared with the approximate 
result of Eq.~(\protect\ref{eq:dLggbW}).}
\label{fig:dLbdep}
\end{figure}

The $b$-dependence is strongly 
influenced by the value of $W$: for $W R \ll \gamma$ (e.g.,
for low mass $e^+e^-$ production) there is an approximate $1/b^2$-
dependence over a considerable range of $b$-values,
 for $W R \approx \gamma$ the value of $x$
is quite small and the dependence is much steeper.
In this case low $b$-values are emphasized anyhow and the effect of 
tagging is less dramatic than for the case of $\rho^0$-production 
with GDR tagging, as discussed above. This was found by Klein in 
\cite{KleinUPC02} for photon-photon processes with nuclear breakup.
In accordance with the present considerations, the effect of tagging is more 
important for low $W$, see his plot as a function of the invariant mass
of a produced muon pair $M_{\mu\mu}$.
A comparison of the luminosity of Eq.~(\ref{eq:dLggbW}) against a full
calculation is shown in Fig.~\ref{fig:dLbdep} for different invariant masses
$W$. In view of the crude approximations made to obtain Eq.~(\ref{eq:dLggbW})
the overall agreement is quite good for small invariant masses,
whereas for the highest invariant masses shown in the plot some part of
the cross section come from the areas beyond the range $u<x^2$. 
This figure also shows the change in the shape of the spectrum for low
and high invariant masses $W$. 

\subsection{Angular correlations}
There can also be angular correlations since the photon polarization
follows the electric field vector of the nuclear fields which in turn
follows $\vec b$ \cite{KleinUPC02}.
For example, the decay $\rho_0 \rightarrow \pi^+ \pi^-$ is sensitive
to the photon polarization, and hence to the direction of $\vec{b}$.
Mutual GDR excitation is another example; the neutron transverse
momentum tends to follow the electric field.  
The amplitude $a_{1,2}(\vec b)$ for mutual excitation is
$$
a_{1,2}(\vec b)=a_1(\vec b) \ a_2(\vec b)
$$
where the amplitude for single GDR excitation $a_i$ ($i=1,2$) is
given, e.g., by Eq.~(3.1.22) of \cite{beba88}.

In intermediate energy heavy ion scattering (in contrast
to the ultrarelativistic case) it is experimentally  
possible to measure directly the momentum transferred to the 
projectile. This determines the impact parameter vector $\vec b$.
The decay products show a dependence on the azimuthal angle 
$\phi$. This dependence has been observed, e.g., in the breakup reaction
$^{11}\mbox{Be} + \mbox{Pb} \rightarrow ^{10}\mbox{Be}+n+\mbox{Pb}$ in
\cite{nak94}, see their Fig.~3. The angular distribution follows a
dipole distribution, see Eq.~(3.1.22) in \cite{beba88}.  For large values of
$\gamma$ the $m=\pm 1$ components dominate and
\begin{equation}
a_{i} \sim \sin\theta \cos\phi,  
\label{eq:afiangle}
\end{equation}
where $\theta$ and $\phi$ are the polar and azimuthal angles of the
emitted neutrons in the system of the nucleus. 

For mutual GDR excitation, the direction of $\vec b$
is the same for both excitations, producing a correlation between the 
emission angles of $\theta_1, \theta_2, \phi_1,$ and $\phi_2$ 
(in the system of the respective emitting nucleus).
The correlation in the relative azimuthal angle $\Delta \phi
= \phi_1 - \phi_2$
of the emitted neutrons is
\begin{equation}
C(\Delta \phi) = 1+ \frac{1}{2} \cos(2\Delta \phi).
\end{equation}

\begin{figure}
\begin{center}
\resizebox{!}{2.5in}{\includegraphics{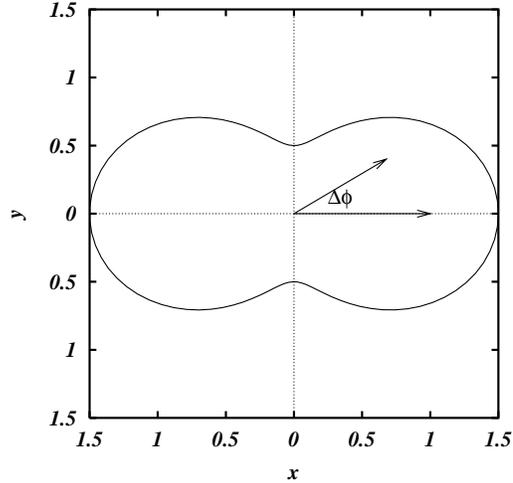}}
\end{center}
\caption{
The angular correlation of two neutrons emitted after mutual GDR excitation
is shown. $\Delta \phi$ denotes the azimuthal angle between the two neutrons}
\label{fig:phicorr}
\end{figure}

This correlation is shown in Fig.~\ref{fig:phicorr}. The neutrons from
the GDR decays can be detected with zero degree calorimeters which are
located downstream of the collision region at relativistic ion
colliders.  At RHIC, the ZDC's are located $\pm 18$ meters 
downstream; they are 
10 cm wide by 18.75 cm high \cite{ZDC}.  The dimensions are limited by
the available space.  Although the sensitivity is reduced near the
edges due to transverse shower leakage, coverage is good for angles
$\theta< 2$ mrad \cite{sebastian}. The excitation energy for a heavy
nucleus GDR is approximately 10~MeV, corresponding to a neutron
momentum of about 140 MeV/c.

For the RHIC neutron longitudinal momentum of 100 GeV, $p_T=140$ MeV/c
corresponds to a deflection of 2.5 cm, well within the ZDC acceptance.
With position sensitive shower-maximum detectors to the ZDC's,
neutron position resolution of order 1 mm should be
achievable \cite{sebastian2}, giving adequate angular resolution to study
neutron-neutron correlations between the ZDCs.  At the LHC, the ZDCs
are located $\pm 116$ meters from the collision regions, giving a
maximum neutron deflection of 5 mm, so determining the neutron angular
direction may be more difficult.

Similar angular correlations will occur for $\rho$ production and
decay to $\pi^+\pi^-$ accompanied by single or mutual GDR excitation.
In its rest frame the $\rho$ decay angular distribution is given by
$|Y_{1M}(\theta, \Phi)|^2$, where $M$ denotes the magnetic quantum
number.  The population of the different $M$-substates depends on the
dynamics of the production process. 

In the heavy ion case the population is relatively simple, since
the photons are almost real, with both helicities equally
populated. It is the simplest assumption that in $\rho$ photoproduction
the $\rho$ has the same
polarization as the initial state photon - this is $s$-channel
helicity conservation (SCHC).  This has been experimentally verified
for photoproduction \cite{fixedtarget,Ballam73} and for
electroproduction \cite{Crittenden97}.  In our case the photons are
linearly polarized, transverse to the beam.  Assuming SCHC, the final
$\rho^0$ state is also linearly polarized, in the same direction as
the equivalent photon.  It is a superposition of $M=\pm 1$ states. The
angular distribution in the rest system is
\begin{equation}
Y_{11} \pm Y_{1-1}= -\sqrt{\frac{3}{\pi}}
\left\{\begin{array}{l}\sin \theta \cos\Phi\\ i \sin \theta \sin\Phi
\end{array}\right.,
\end{equation}
quite analogous to Eq.~(\ref{eq:afiangle}) for the GDR decay.  For
exclusive $\rho$ production, the direction of the impact parameter is
unknown, and one has to integrate over all impact parameters. However,
if the coherent $\rho$-production is 
accompanied by a GDR excitation, the azimuthal angles of the neutron
and the $\pi^+ \pi^-$-pair would be correlated, as with mutual GDR
excitation. The GDR decay provides a way to measure the photon polarization.
The same correlation would be observed between two
$\rho^0$ that are produced together \cite{skjn99}.

In $e^+e^-$ pair production one also expects a dependence on
the azimuthal angle $\phi$ which is defined by the impact parameter
vector $\vec b$. For polarized (real) photons this can be seen from the 
analysis of, e.g., \cite{McMaster61}. 
For linearly polarized photons the electron is most likely
to be emitted in the plane of the polarization \cite{McMaster61}.
It will be of interest to do such theoretical calculations in more detail
and work on this is currently in progress.
E.g. in two-pair production one might expect to see such correlations, 
which leave their trace on the impact parameter of the collision.
The formula for one-pair production in lowest order semiclassical 
approximation is given e.g. in \cite{LandauLQED,HenckenTB95b,Alscher97}. This
formula contains the correlations of the impact parameter
and the momenta of the outgoing $e^+ e^-$ pair.

We conclude this section with a general remark on multiphoton processes.
The observation of two electromagnetic processes with 
photon energies $\omega_1$ and $\omega_2$ gives information
about the range of impact parameters of the collision. One has
\begin{equation}
P_{12}(b)= P_1(b) P_2(b)= N(\omega_1,b) N(\omega_2,b) 
\sigma_{\gamma}(\omega_1) \sigma_{\gamma}(\omega_2)
\end{equation}
We use a simple approximation for the 
$b$-dependent equivalent photon spectrum $N(\omega,b)$:
\begin{equation}
  N(\omega,b)= \frac{Z_1^2 \alpha}{\pi^2 b^2}\quad\mbox{for
$2R< b < \frac{\gamma}{\omega}$}
\end{equation}
Thus such an interaction takes place in the b-range of 
$2R<b< \frac{\gamma}{\omega_{max}}$, where $\omega_{max}$ is 
the larger of $\omega_1$ and $\omega_2$. 

\section{Conclusions}
\label{sec:conclusions}

Due to the strong fields of highly charged ions multiphoton processes
are abundant in UPCs. The important modes were identified. 
Quite generally, the excitation of the different modes
occurs independently and the corresponding amplitudes can be
factorized.  They do not influence each other. This is essentially due
to the very high energy of the heavy ions, which is much higher than
the energy $\omega$ of the equivalent photon.  Only a small fraction
of the total kinetic energy of the nuclei is lost in the UPC. Due to
the coherence condition one has $x_{max}=\frac{\omega}{E}=\frac{1}{R
M_N A} =\frac{\lambda_C (A)}{R} \ll 1$.  $M_N$ is the nucleon mass,
and $\lambda_C(A)$ is the Compton wavelength of the nucleus.

The main result of this paper may seem intuitively obvious.
However, the experimental implications are significant.
The factorization assumption was used in detailed
theoretical calculations  of $\rho$-production and GDR
excitation in \cite{BaltzKN02b} (following Table 2 it is said there
that ``the validity of the assumption of factorization is 
hard to prove rigorously \dots''), and \cite{BaltzKN02}
and in the corresponding experimental analysis \cite{Adleretal}.
It is important to show also how this factorization 
arises in the theoretical formulation of the process.
Now there is a formal basis to discuss these
questions theoretically. 

Ultraperipheral heavy ion collisions provide a strong 
source of equivalent photons up to very high energies.
This offers the unique possibility to study 
photon-hadron (nucleus) and photon-photon processes
in hitherto inaccessible regions, see \cite{BaurHTSK}.
Many UPC processes like GDR excitation are now recognized as
useful for practical matters like luminosity measurement and
impact-parameter dependent triggers. In this 
case, as was already shown explicitly in \cite{BaltzKN02}
one emphasizes the hard part of the spectrum, due to the selection
of low $b$ values. 

\section*{Acknowledgments}

This work was partially supported by the US Dept. of Energy, under
contract No DE-AC-03076SF00098. We would like to thank Sebastian White 
for interesting discussions on this subject.
We thank also the  Institute for Nuclear Theory at
the University of Washington for its hospitality and the Department of Energy
for partial support and the organizers of the program ``The first three 
years of heavy ion physics at RHIC''. The  ``UPC week'' of this program,
which took place April 21st--27th 2003, 
was vital for this paper, both due to the interaction 
of some of the present authors and also to the stimulating
atmosphere created by all the participants.

\end{document}